\documentclass[conference]{IEEEtran}

\usepackage{epsfig,epsf,amsmath,amssymb,amsthm,scalefnt,multirow,dsfont,subfig}
\usepackage{xcolor}
\usepackage{float}
\usepackage{cite}
\usepackage{graphicx}

\newcommand{\argmax}{\mathop{\mathrm{argmax}}}
\newcommand{\argmin}{\mathop{\mathrm{argmin}}}

  \def\cC{{\mathcal{C}}} 
 \def\cF{{\mathcal{F}}}  
   
 \def\cN{{\mathcal{N}}}

\def\argmin{\mathop{\mathrm{argmin}}}
\def\argmax{\mathop{\mathrm{argmax}}}

\def\b0{{\pmb{0}}} 

\def\ba{{\mathbf{a}}}  \def\bc{{\mathbf{c}}} 
 \def\bff{{\mathbf{f}}} \def\bg{{\mathbf{g}}} \def\bh{{\mathbf{h}}}

  \def\bw{{\mathbf{w}}}

\def\bA{{\mathbf{A}}}   
   \def\bH{{\mathbf{H}}}

 \def\bR{{\mathbf{R}}}  
  \def\bW{{\mathbf{W}}}

\begin{document}

\title{Exploiting the Preferred Domain of FDD Massive MIMO Systems with Uniform Planar Arrays}

\author{\IEEEauthorblockN{Junil Choi$^{\dag}$, Taeyoung Kim$^{\ddag}$, David J. Love$^{\dag}$, and Ji-yun Seol$^{\ddag}$}\\
\IEEEauthorblockA{$^{\dag}$School of Electrical and Computer
Engineering, Purdue University, West Lafayette, IN\\
Email: choi215@purdue.edu, djlove@purdue.edu\\
$^{\ddag}$Communications Research Team (CRT), DMC R\&D Center, Samsung Electronics Co., Ltd., Suwon, Korea\\
Email: ty33.kim@samsung.com, jiyun.seol@samsung.com}} \maketitle

\begin{abstract}
Massive multiple-input multiple-output (MIMO) systems hold the potential to be an enabling technology for 5G cellular.  Uniform planar array (UPA) antenna structures are a focus of much commercial discussion because of their ability to enable a large number of antennas in a relatively small area.  With UPA antenna structures, the base station can control the beam direction in both the horizontal and vertical domains simultaneously.  However, channel conditions may dictate that one dimension requires higher channel state information (CSI) accuracy than the other.  We propose the use of an additional one bit of feedback information sent from the user to the base station to indicate the preferred domain on top of the feedback overhead of CSI quantization in frequency division duplexing (FDD) massive MIMO systems.  Combined with variable-rate CSI quantization schemes, the numerical studies show that the additional one bit of feedback can increase the quality of CSI significantly for UPA antenna structures.
\end{abstract}

\section{Introduction}\label{sec1}
Future wireless cellular systems are expected to deploy a large number of antennas, e.g., 10s-100s of antennas, at the base station \cite{massive_mimo5,fdmimo,massive_mimo6}.  This trend of deploying a large number of antennas is now well known as massive multiple-input multiple-output (MIMO) systems.  After pioneering work done by Marzetta in \cite{massive_mimo1}, many follow up works have been dedicated to verifying the benefits, potential uses, drawbacks, and limitations of massive MIMO systems.  We refer to \cite{massive_mimo5,massive_mimo6} and references therein for details.

Most of the previous works on massive MIMO focused on time division duplexing (TDD) to circumvent downlink channel state information (CSI) estimation and quantization problems.  Frequency division duplexing (FDD) is extremely challenging to implement with massive MIMO because most previous solutions for CSI estimation and feedback design become impractical as the number of antennas grows large.  However, most current wireless cellular systems are based on FDD, and backward compatibility is crucial for advanced wireless communication technologies.  Thus, it is of great interest to solve CSI estimation and quantization problems for massive MIMO systems.

There are some existing works dedicated to making FDD massive MIMO practical.  Low complexity CSI quantization schemes are developed in \cite{CK,upa_codebook,jsdm,tcom_ntcq,dawei,tec} where \cite{tec} specifically focused on backward compatibility with the 3GPP LTE-Advanced standard.  The schemes in \cite{CK,tcom_ntcq,tec} are variable-rate CSI quantization techniques that adaptively control the feedback overhead of CSI quantization.  For the channel sounding problem, \cite{song,cl_training_jstsp} showed that training overhead can be significantly reduced by adapting the training signals using knowledge of the long-term channel statistics.  These past works, however, have not considered practical antenna structures of massive MIMO systems with a few exceptions in \cite{upa_codebook,dawei} that consider uniform planar arrays (UPAs), also sometimes referred to as uniform rectangular arrays (URAs).

Interest in UPAs for massive MIMO deployment is growing, mainly because of the ability of a UPA to house a large number of antennas in a small area.  Massive MIMO with a UPA is sometimes referred to as three-dimensional (3D) MIMO \cite{3dmimo1,3dmimo2,3dmimo3} because the base station can control the beam direction in both the horizontal and vertical domains simultaneously using the two-dimensional structure of a UPA.  In this case, the channel might require more CSI accuracy for the horizontal \textit{or} vertical domain depending on the scenario.  Thus, we propose to use one additional bit of feedback to achieve the full benefit of variable-rate CSI quantization schemes in UPA scenarios.

The additional one bit of feedback explicitly indicates the \textit{preferred domain} between the horizontal and vertical domains.  It is important to point out that the concept of the preferred domain in CSI quantization is novel because current wireless communication standards such as 3GPP LTE and LTE-Advanced only focus on one-dimensional antenna array structures \cite{lte}.  With one-dimensional antenna array structures such as a uniform linear array or dual-polarized linear array, the user only sees the horizontal domain beam pattern.  The key idea of the proposed scheme is that the user re-indexes the channel elements to be quantized depending on the domain that needs more precise quantization.  Simulation results show that the additional one bit of feedback with appropriate CSI quantization schemes can increase the quality of CSI significantly for massive MIMO with UPA antenna structures.

The paper is organized as follows.  We explain our system model and discuss CSI quantization techniques for UPA structures in Section \ref{sec2}.  In Section \ref{sec3}, we propose our one bit of additional feedback idea.  Numerical studies that evaluate the proposed idea are shown in Section \ref{sec4}, and conclusions follow in Section \ref{sec5}.

\section{System Model and CSI Quantization}\label{sec2}
We describe our system model first.  Then, we explain problems and possible solutions of CSI quantization for UPA antenna structures.

\subsection{System Model}
\begin{figure}[t]
  \centering
  \includegraphics[scale = 0.45]{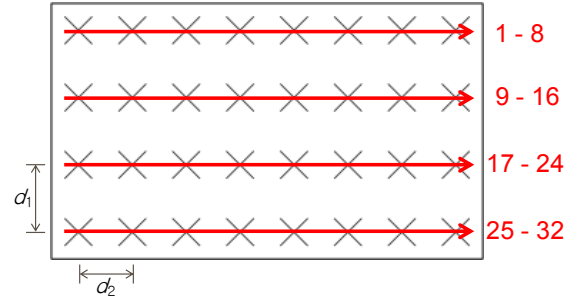}\\
  \caption{An example of UPA antenna structure of $N_v=4$ and $N_h=8$ with antenna element indexing.}\label{upa1}
\end{figure}

We consider a multiuser (MU) MIMO block-fading channel where the base station with $N_p = N_v \times N_h$ antennas serves $K$ single antenna users simultaneously.  The base station is equipped with $N_v$ rows and $N_h$ columns of UPA antennas with antenna spacing $d_1$ for the vertical domain and $d_2$ for the horizontal domain.  We define the $(k,l)$-th antenna as the antenna element in the $k$-th row and the $l$-th column of a UPA, which results in the $(k,l)$-th antenna corresponding to the $l+N_h(k-1)$-th element of $\bh[m]$. Fig. \ref{upa1} shows an example of $4\times 8$ UPA structure with antenna indexing.

Assuming equal power allocation, the received signal of the user $i$ at the $m$-th fading block is written as
\begin{align*}
  y_{i}[m]&=\sqrt{\frac{\rho}{K}}\bh_{i}^H[m] \bw_{i}[m] s_{i}[m]  \\
  &\qquad +\sqrt{\frac{\rho}{K}}\sum_{u=1,u\neq i}^{K} \bh_{i}^H[m]\bw_{u}[m] s_{u}[m]+ z_{i}[m]
\end{align*}
where $\bh_{i}[m]\in \mathbb{C}^{N_p}$, $\bw_{i}[m]\in \mathbb{C}^{N_p}$, and $z_i[m] \sim \cC\cN(0,1)$ are the channel vector, the unit norm beamforming vector, and the complex additive white Gaussian noise of the $i$-th user, respectively.  $s_{i}[m]$ is the data signal of the $i$-th user with $E\left[s_i\left[k\right]\right] = 0$ and $E\left[|s_i[m]|^2\right] = 1$, and $\rho$ is the transmit signal-to-noise ratio (SNR).

Note that $\bw_{i}[m]$ is a function of the quantized CSI received from all users, i.e., $\hat{\bh}_{i}[m]$ for $i=1,\ldots,K$.  We rely on well-known zeroforcing beamforming (ZFBF) for $\bw_{i}[m]$ \cite{zfbf}.  Let $\hat{\bH}[m]$ be the composite matrix of $\hat{\bh}_{i}$ as
\begin{equation}
  \hat{\bH}[m] = \left[\hat{\bh}_{1}[m]~\cdots~\hat{\bh}_{K}[m]\right].\label{composite_H}
\end{equation}
Then, the composite ZFBF precoding matrix $\bW[m]$ is given as
\begin{equation*}
  \bW[m] = \hat{\bH}[m]\left(\hat{\bH}[m]^H\hat{\bH}[m]\right)^{-1},
\end{equation*}
and the beamforming vector for user $i$ becomes
\begin{equation*}
\bw_{i}[m]=\bW_{(:,i)}[m]
\end{equation*}
where $\bA_{(:,k)}$ is the $k$-th column of the matrix $\bA$.

The instantaneous signal-to-interference-noise ratio (SINR) of the $i$-th user is given as
\begin{equation*}
\mathrm{SINR}_{i}[m] = \frac{\left|\bh_{i}^H[m] \bw_{i}[m]\right|^2}{\sum\limits_{u=1,u\neq i}^{K}\left|\bh_{i}^H[m] \bw_{u}[m]\right|^2+\frac{K}{\rho}},
\end{equation*}
and the corresponding sum-rate of the $m$-th fading block is written as
\begin{equation*}
R_{sum}[m] = \sum_{u=1}^{K}\log_2\left(1+\mathrm{SINR}_{u}[m]\right)
\end{equation*}
assuming Gaussian signaling.

We consider spatially and temporally correlated channels in this paper.  Thus, we model the channel vector of the $i$-th user as
\begin{equation}\label{channel_model}
  \bh_{i}[m] = \eta_{i} \bh_{i}[m-1]+\sqrt{1-\eta_{i}^2}\bR_{i}^{\frac{1}{2}}\bg_{i}[m]
\end{equation}
where $0\leq \eta_{i} \leq 1$ is the temporal correlation coefficient, $\bR_{i} = E\left[\bh_{i}[m]\bh_{i}[m]^H\right]$ is a spatial correlation matrix, and $\bg_{i}[m] \sim \cC\cN(0,1)$ is an innovation process with i.i.d. Rayleigh fading components.  At $m=0$, we have $\bh_{i}[0]=\bR_{i}^{\frac{1}{2}}\bg_{i}[0]$.

For the spatial correlation matrix $\bR_{i}$, we adopt the UPA spatial correlation model from \cite{dawei}, where the correlation between the $(k,l)$-th and $(p,q)$-th antenna elements is given as
\begin{equation}\label{R_model}
  [\bR_{i}]_{(k,l),(p,q)}=\frac{D_1^{i}}{\sqrt{D_5^{i}}}e^{-\frac{D_7^{i}+(D_2^{i}(\sin \phi_{i})\sigma_{i})^2}{2D_5^{i}}}e^{j\frac{D_2^{i} D_6^{i}}{D_5^{i}}}
\end{equation}
with the variables
\begin{align*}
  D_1^{i}&=e^{j\frac{2\pi d_1}{\lambda}(p-k)\cos \theta_{i}}e^{-\frac{1}{2}(\xi_{i} \frac{2\pi d_1}{\lambda})^2(p-k)^2\sin^2\theta_{i}},\\
  D_2^{i}&=\frac{2\pi d_2}{\lambda}(q-l)\sin\theta_{i},\\
  D_3^{i}&=\xi_{i}\frac{2\pi d_2}{\lambda}(q-l)\cos\theta_{i},\\
  D_4^{i}&=\frac{1}{2}\left(\xi_{i}\frac{2\pi}{\lambda}\right)^2d_1 d_2 (p-k)(q-l)\sin(2\theta_{i}),\\
  D_5^{i}&= \left(D_3^{i}\right)^2 ((\sin \phi_{i})\sigma_{i})^2+1,\\
  D_6^{i}&= D_4^{i} ((\sin \phi_{i})\sigma_{i})^2+\cos\phi_{i},\\
  D_7^{i}&= \left(D_3{i}\right)^2\cos^2\phi_{i} - \left(D_4{i}\right)^2 ((\sin \phi_{i})\sigma_{i})^2-2D_4^{i}\cos\phi_{i}
\end{align*}
where $\lambda$ is the carrier frequency wavelength, $\phi_{i}$ is the mean horizontal angle-of-departure (AoD), $\theta_{i}$ is the mean vertical AoD, $\sigma_{i}$ is the standard deviation of horizontal AoD, and $\xi_{i}$ is the standard deviation of vertical AoD of the $i$-th user.  The boresight angle of $\phi_{i}$ and $\theta_{i}$ are both $\pi/2$.
\begin{figure*}
\begin{align}
\nonumber \argmax_{\bc\in\mathcal{C}}|\bh^H[m]\bc|^2 &=\argmin_{\bc\in\mathcal{C}}\min_{\psi\in[0,2\pi)}\left\lVert\bh[m]-e^{j\psi}\bc\right\rVert^2\\
&= \argmin_{\bc\in\mathcal{C}}\min_{\psi\in[0,2\pi)}\sum_{n=1}^{N_p/L}\left\lVert \bh_{[L(n-1)+1:Ln]}[m]-e^{j\psi}\bc_{[L(n-1)+1:Ln]}\right\rVert^2.\tag{4} \label{tec_crit}
\end{align}
\hrule
\end{figure*}

\subsection{CSI Quantization for UPA}\label{ura_csi}
Because CSI quantization is performed independently at each user, we drop the user index $i$ in this subsection.  To obtain a reasonable CSI quality, we assume that the total CSI quantization overhead $B_{\mathrm{CSI}}$ is given as
\begin{equation*}
  B_{\mathrm{CSI}}=BN_p
\end{equation*}
where $B$ is the quantization bits per antenna element.  If we rely on the conventional approach of using $N_p$-dimensional vector quantized codebooks for CSI quantization, the computational complexity that grows exponentially with $B_{\mathrm{CSI}}$ becomes a serious problem for large $N_p$ \cite{CK}, which is typically the case of UPA antenna structures.

There are several ways to solve the CSI quantization complexity issue for UPA antenna structures. It was shown in \cite{dawei} that the spatial correlation matrix $\bR$ can be approximated as\footnote{In \cite{dawei}, the approximation is given as $\bR \approx \bR_h \otimes \bR_v$ because \cite{dawei} indexed antenna elements vertically.}
\begin{equation} \tag{3}\label{R_approx}
  \bR \approx \bR_v \otimes \bR_h
\end{equation}
where $\otimes$ is the Kronecker product and
\begin{align*}
[\bR_v]_{k,p} &= e^{-\frac{1}{2}\left(\xi\frac{2\pi d_1}{\lambda}\right)^2(p-k)^2\sin^2 \theta} e^{j\frac{2\pi d_1}{\lambda}(p-k)\cos \theta},\\
[\bR_h]_{l,q} &= \frac{1}{\sqrt{D_5}}e^{-\frac{D_3^2\cos^2\phi + (D_2(\sin \phi)\sigma)^2}{2D_5}}e^{j\frac{D2 \cos \phi}{D_5}}
\end{align*}
are the spatial correlation matrices of the vertical and horizontal domains, respectively.

The approximation in \eqref{R_approx} suggests to quantize the CSI of horizontal and vertical domains separately.  Let $\bar{\bH}[m]$ be the matrix consists of the elements of $\bh[m]$ as\footnote{The $(k,l)$-th element of $\bar{\bH}[m]$ corresponds to the $(k,l)$-th antenna element of the UPA antenna structure.}
\begin{equation*}
  \bar{\bH}[m] = \begin{bmatrix}\bh_{[1:N_h]}^T[m] \\ \bh_{[N_h+1:2N_h]}^T[m] \\ \vdots \\ \bh_{[N_v(N_h-1)+1:N_vN_h]}^T[m]\end{bmatrix}
\end{equation*}
where $\ba_{[n_1:n_2]}$ is the sub-vector of $\ba$ consists of the $n_1$-th to $n_2$-th entries.  Then the user can quantize the CSI of horizontal and vertical domains as
\begin{align*}
  \bc_{h,\mathrm{opt}}[m] & = \argmax_{\bc_h\in \cC_h}\left\Vert \bar{\bH}[m] \bc_h \right\Vert^2, \\
  \bc_{v,\mathrm{opt}}[m] & = \argmax_{\bc_v\in \cC_v}\left\Vert \bc_v^H \bar{\bH}[m] \right\Vert^2
\end{align*}
with $B_h$-bits, $N_h$-dimensional codebook $\cC_h$ and $B_v$-bits, $N_v$-dimensional codebook $\cC_v$ with a constraint $B_h+B_v=B_{\mathrm{CSI}}$ (or $B_{\mathrm{CSI}}+1$ taking the proposed one bit of additional feedback into account). Once the user feeds back $\bc_{h,\mathrm{opt}}[m]$ and $\bc_{v,\mathrm{opt}}[m]$, the base station can reconstruct the quantized CSI as
\begin{equation*}
\hat{\bh}[m] = \bc_{v,\mathrm{opt}}[m] \otimes \bc^*_{h,\mathrm{opt}}[m].
\end{equation*}
We dub this approach the \textit{Kronecker-product approach} and consider it as the baseline of performance evaluation.\footnote{Although not exactly the same, similar CSI quantization techniques as the Kronecker-product approach have been proposed in \cite{dawei,upa_codebook}.}

On the other hand, we can quantize $\bh[m]$ in a block-wise manner.  For example, we can quantize $L\ll N_p$ dimensional sub-channel vectors separately using $BL$ bits for each sub-channel vector.\footnote{Note that $L$ is the system parameter that needs to be optimized to minimize CSI quantization error with fixed $B_{\mathrm{CSI}}$, which is out of scope of this paper.}  The problem is that the codeword selection criterion of using an $N_p$-dimensional codebook cannot be decomposed into $N_p/L$ sub-problems because \cite{CK}
\begin{align*}
  &\argmax_{\bc\in\mathcal{C}}|\bh^H[m]\bc|^2\\
  &\qquad \neq \argmax_{\bc\in\mathcal{C}}\sum_{n=1}^{N_p/L}|\bh_{[L(n-1)+1:Ln]}^H[m]\bc_{[L(n-1)+1:Ln]}|^2
\end{align*}
assuming $L$ divides $N_p$.  However, we can transform the problem as in \eqref{tec_crit}.  Thus, we can quantize each separate block of $\bh[m]$ with a small number of $BL$ bits in a noncoherent manner,\footnote{Note that $\psi$ is only an auxiliary variable during the optimization process, and the user does not need to feed back the information of $\psi$ to the base station.  We refer to \cite{tec} for details.} which can reduce the complexity significantly.

Taking this transformed problem into account, the trellis-extended codebook (TEC) and trellis-extended successive phase adjustments (TE-SPA) that quantize $\bh[m]$ in a block-wise manner have been proposed in \cite{tec}.  To be specific, TEC is a base codebook that quantizes $\bh[0]$, and TE-SPA is a differential codebook \cite{tm_correlated1,tm_correlated2,tm_correlated8,tm_correlated4,tm_correlated5,tm_correlated6,tm_correlated9} that exploits the temporal correlation of channels and quantizes $\bh[m]$ for $m\geq 1$ using the previously quantized CSI $\hat{\bh}[m-1]$.  Another advantage of TEC and TE-SPA is that they support variable-rate CSI quantization, which makes it easy to adapt feedback overhead depending on requirements.  In this paper, we assume the one bit of additional feedback idea that will be explained in the next section is combined with TEC and TE-SPA.  However, the proposed one bit of additional feedback idea can be applied to any kind of block-wise CSI quantization schemes.

\section{One Bit of Additional Feedback}\label{sec3}
Note that a UPA is a two-dimensional antenna structure and can control the beam direction not only of the vertical domain but also of the horizontal domain.  Moreover, the approximation in \eqref{R_approx} shows that the CSI of one domain might need more accurate quantization than the other.  Thus, we propose to have one bit of additional feedback, which indicates the preferred domain, from the user to the base station.

We drop the user index $i$ and the fading block index $m$ to simplify notation.  We assume $N_v$ and $N_h$ are multiples of $L$.  We further assume that $\bh$ is quantized in a block-wise manner with the block size $L$.  Using the $L$-dimensional codebook $\mathcal{F}$, we first generate a candidate quantized CSI $\hat{\bh}_1$ as
\begin{equation}
\hat{\bh}_1=\begin{bmatrix}\bff_{1,1}^T\cdots \bff_{1,N_p/L}^T\end{bmatrix}^T,\label{TEC_quant}
\end{equation}
where
\setcounter{equation}{4}
\begin{align*}
&\left(\bff_{1,1},\ldots,\bff_{1,N_p/L}\right) = \\
&\qquad \argmin_{\bff_{1,n} \in \mathcal{F}}\min_{\psi\in[0,2\pi)}\sum_{n=1}^{N_p/L}\left\lVert \bh_{[L(n-1)+1:Ln]}-e^{j\psi}\bff_{1,n}\right\rVert^2.
\end{align*}
This problem can be solved by TEC, TE-SPA\footnote{Using a trellis structure, we can have a codebook expansion effect in TEC and TE-SPA, e.g., 8 codewords are available for $\cF$ while only allocating 2 bits per sub-problem in \eqref{TEC_quant}.  We refer to \cite{tec} for more details.}, or any other block-wise CSI quantization schemes.  We assume the elements of $\mathcal{F}$ are properly normalized to have $\|\hat{\bh}_1\|^2=1$.

In addition to $\hat{\bh}_1$ in \eqref{TEC_quant}, we generate another candidate quantized CSI $\hat{\bh}_2$ as
\begin{equation}
\hat{\bh}_2=\begin{bmatrix}\bff_{2,1}^T\cdots \bff_{2,N_p/L}^T\end{bmatrix}^T,\label{TEC_rearranged_quant}
\end{equation}
with
\begin{align*}
&\left(\bff_{2,1},\ldots,\bff_{2,N_p/L}\right) = \\
&\qquad \argmin_{\bff_{2,n} \in \mathcal{F}}\min_{\psi\in[0,2\pi)}\sum_{n=1}^{N_p/L}\left\lVert \widetilde{\bh}_{[L(n-1)+1:Ln]}-e^{j\psi}\bff_{2,n}\right\rVert^2
\end{align*}
where $\widetilde{\bh}$ is the rearranged channel vector of $\bh$.  To explicitly describe $\widetilde{\bh}$, we define two functions
\begin{equation*}
  c(n) \triangleq \left\lceil\frac{nL}{N_v} \right\rceil,
\end{equation*}
\begin{equation*}
  m(n) \triangleq \left((n-1)~\text{mod}~\frac{N_v}{L}\right).
\end{equation*}
With these functions, $\widetilde{\bh}$ is given as
\begin{equation*}
  \widetilde{\bh}=\begin{bmatrix}\check{\bh}_1^T \cdots \check{\bh}_{N_p/L}^T\end{bmatrix}^T
\end{equation*}
where
\begin{align*}
  \check{\bh}_n = &\left[h_{c(n)+Lm(n)N_h},h_{c(n)+(Lm(n)+1)N_h},\right.\\
&\qquad \qquad \qquad \left.\cdots,h_{c(n)+(L(m(n)+1)-1)N_h}\right]^T
\end{align*}
for $n=1,\ldots,N_p/L$.  The final quantized CSI $\hat{\bh}$ is then given as
\begin{equation*}
  \hat{\bh} = \begin{cases}\hat{\bh}_1\quad \text{if}~~|\bh^H\hat{\bh}_1|^2\geq |\widetilde{\bh}^H\hat{\bh}_2|^2\\
  \hat{\bh}_2\quad \text{if}~~|\bh^H\hat{\bh}_1|^2< |\widetilde{\bh}^H\hat{\bh}_2|^2\end{cases}.
\end{equation*}

\begin{figure}[t]
\centering
\subfloat[The channel vector $\bh$ is quantized with $\hat{\bh}_1$.]{
\includegraphics[width=0.6\columnwidth]{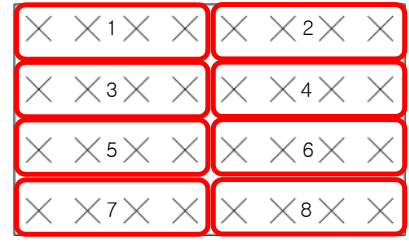}
\label{horizon}
}\\
\begingroup
\captionsetup[subfigure]{width=5in}
\subfloat[The rearranged channel vector $\widetilde{\bh}$ is quantized with $\hat{\bh}_2$.]{
\includegraphics[width=0.6\columnwidth]{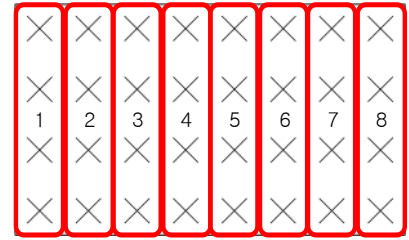}
\label{vertical}
}
\endgroup
\caption{The conceptual explanation of the proposed idea with $4\times 8$ UPA antenna structure and $L=4$.  The red box is quantized with $\bff_{1,n}$ in $(a)$ and $\bff_{2,n}$ in $(b)$ for $n=1,\ldots,N_p/L$ where the numbers in the red boxes correspond to $n$.}
\label{domain_selection}
\end{figure}

Note that we can adopt an arbitrary $L$-dimensional codebook $\mathcal{F}$ when we solve \eqref{TEC_quant} and \eqref{TEC_rearranged_quant}, and it is well known that standard codebooks such as DFT or LTE codebooks have good beam directivity when antenna spacing is small.  Thus, it would be a good choice to adopt DFT or LTE codebooks as $\mathcal{F}$ for block-wise CSI quantization schemes for UPA antenna structures.

Now, we conceptually explain the proposed idea using\footnote{A figure similar to Fig. \ref{domain_selection} was shown in \cite{alu}; however, the figure in \cite{alu} was to propose downlink training design instead of CSI feedback.  The concept of the preferred domain for CSI feedback in this paper is new.} Fig.~\ref{domain_selection}.  It might be beneficial for the user to quantize one domain more precisely than the other depending on the scenario, e.g., user location or user movement.  If the user needs to quantize the horizontal domain more precisely, than the user should quantize the channel \textit{horizontally} using structured codewords.  On the other hand, the user needs to quantize the channel \textit{vertically} with structured codewords if the user needs more accurate CSI for the vertical domain.  With these observations, the candidate CSI $\hat{\bh}_1$ quantizes the horizontal domain of the channel more precisely as in (a) of Fig. \ref{domain_selection}, while $\hat{\bh}_2$ quantizes the vertical domain of the channel more accurately as in (b) of Fig. \ref{domain_selection}.  By comparing $\hat{\bh}_1$ and $\hat{\bh}_2$, the user can determine the more important domain for the user and select more accurately quantized CSI.  The additional one bit of feedback indicates which candidate CSI is selected between $\hat{\bh}_1$ and $\hat{\bh}_2$.

\noindent \textbf{Remark 1:} If we rely on the conventional approach of using $N_p$-dimensional vector quantized codebook for CSI quantization, then the proposed one bit of additional feedback has eventually the same effect of using a $B_{\mathrm{CSI}}+1$-bits vector quantized codebook.  However, block-wise CSI quantization schemes are not able to exploit additional few bits (one bit in our propopsed idea) of feedback in this manner.  Thus, the proposed idea is particularly suitable to block-wise CSI quantization schemes.

\noindent \textbf{Remark 2:} The channel vector rearrangement should be dependent on antenna indexing and block-wise CSI quantization schemes.  However, the proposed idea can be applied to any scenario without difficulty.

\section{Simulation Results}\label{sec4}
We evaluate the proposed idea with Monte-Carlo simulations in this section.  We assume the base station serves $K=10$ users simultaneously with ZFBF precoding.\footnote{To perform ZFBF precoding properly, we only consider the channel realizations when the composite quantized channel matrix $\hat{\bH}[m]$ given in \eqref{composite_H} is full rank.  This can be thought as a very simple scheduling scheme.}  We adopt the channel model in \eqref{channel_model} with the spatial correlation matrix in \eqref{R_model}.  Considering a practical UPA structure and cell layout, we set the vertical antennna spacing to $d_1=0.9\lambda$, horizontal antenna spacing to $d_2=0.5\lambda$, randomly uniformly generate the mean horizontal AoD $\phi_{i}$ in $[\pi/6,5\pi/6)$, the mean vertical AoD $\theta_{i}$ in $[\pi/12,\pi/3)$, and the standard deviations of horizontal AoD $\sigma_{i}$ and vertical AoD $\xi_{i}$ both in $[\pi/18,\pi/12)$ independently for all users in each channel realization.  The transmit SNR $\rho$ is set to 10dB for all simulations.

We first compare the MU-MIMO sum-rate of three schemes, i.e., the Kronecker-product approach explained in Section \ref{ura_csi}, TEC from \cite{tec}, and TEC with the proposed one bit of additional feedback.  We set $L=4$ and adopt the $L$-dimensional DFT codebook that satisfies $B=1/2$ bit per antenna quantization for TEC.  The Kronecker-product approach is based on two $N_v$ and $N_h$-dimensional DFT codebooks of sizes $B_v=\frac{1}{4}N_p$ and $B_h=\frac{1}{4}N_p+1$ bits.  The feedback overhead of TEC is given as $B_{\mathrm{CSI}}=\frac{1}{2}N_p$ while that of the other two schemes is $B_{\mathrm{CSI}}+1$.  We only consider the $m=0$-th fading block in this simulation because all three schemes do not exploit any temporal correlation, .
\begin{figure}[t]
\centering
\includegraphics[width=1\columnwidth]{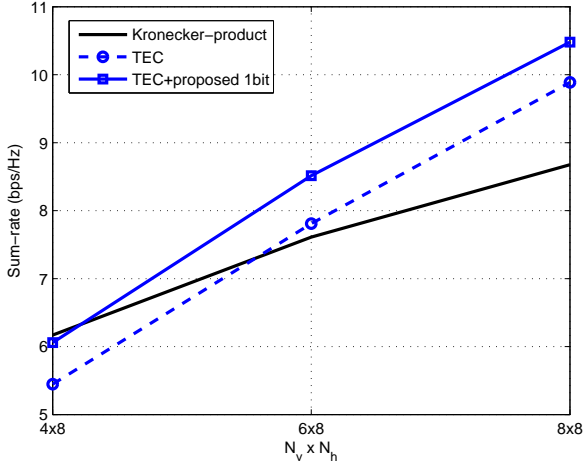}
\caption{Sum-rate (bps/Hz) according to $N_p=N_v \times N_h$ antennas.  The feedback overhead of TEC from \cite{tec} is $B_{\mathrm{CSI}}=\frac{1}{2}N_p$ bits while that of other two schemes is $B_{\mathrm{CSI}}+1$ bits.}
\label{oneshot_res}
\end{figure}

In Fig. \ref{oneshot_res}, we plot the sum-rates of the three schemes according to different combinations of $N_p=N_v \times N_h$ UPA antennas.  It is clear that the proposed one bit of additional feedback gives around 0.5 to 0.8 bps/Hz gain of sum-rate depending on antenna configurations.  Moreover, TEC with the proposed one bit of additional feedback is comparable to the Kronecker-product approach in the $4\times 8$ UPA antenna scenario and outperforms the Kronecker-product approach in the $6\times 8$ and $8\times 8$ UPA antenna scenarios.  Although we could not simulate higher dimensions of UPA because of the complexity issue of the Kronecker-product approach (e.g., a 20 bit codebook is needed for the Kronecker-product approach when $N_p=10\times 8$), we expect that TEC with the proposed one bit of additional feedback would keep outperforming the Kronecker-product approach in larger UPA dimensions as well.
\begin{figure}[t]
\centering
\includegraphics[width=1\columnwidth]{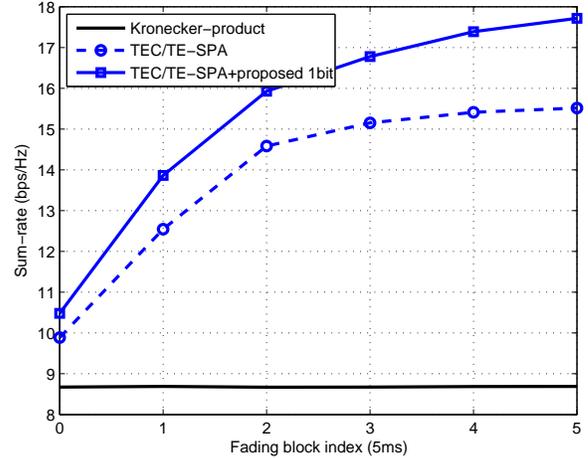}
\caption{Sum-rate (bps/Hz) according to the fading block index $m$ with $8\times 8$ UPA antennas.  TEC quantizes CSI at $m=0$ and TE-SPA quantizes CSI at $m\geq 1$.  The feedback overheads of all three schemes are the same as in Fig. \ref{oneshot_res} for all $m$.}
\label{time_res}
\end{figure}

If Fig. \ref{time_res}, we plot the sum-rates of three schemes according to the fading block index $m$ with $8\times 8$ UPA antennas.  We adopt Jakes' model to generate the temporal correlation coefficient as \cite{prok}
\begin{equation*}
  \eta = J_0\left(2\pi f_D \tau\right)
\end{equation*}
where $J_0(\cdot)$ is the zeroth-order Bessel function, $f_D$ is the Doppler frequency, and $\tau$ is the channel instantaneous interval.  With 2.5GHz carrier frequency, 3km/h user velocity, and 5ms channel instantiation interval, we have $\eta=0.9881$.  Note that the Kronecker-product approach does not exploit temporal correlation of channels while other two schemes exploit the temporal correlation by quantizing CSI with TE-SPA \cite{tec} for $m\geq 1$.

The figures clearly show that the two block-wise CSI quantization schemes outperform the Kronecker-product approach as $m$ increases.  We might be able to take advantage of the temporal correlation even for the Kronecker-product approach; however, to the best of our knowledge, no such scheme has been developed yet.  Note that the proposed one bit of additional feedback significantly increases CSI quality for all $m$, which verifies its effectiveness.

\section{Conclusion}\label{sec5}
We proposed to have one bit of additional feedback on top of CSI quantization overhead from the user to the base station for FDD massive MIMO systems in this paper.  The proposed one bit of additional feedback indicates the preferred domain of the user's channel, which is a new concept in CSI quantization.  This concept has become available because of the use of UPA antenna structures that are getting more interest due to massive MIMO systems.  The proposed idea is well suitable to block-wise CSI quantization schemes which are shown to be appropriate for UPA antenna structures.  The simulation results showed that the proposed one bit of additional feedback idea can significantly increase the quality of quantized CSI in practical UPA channel conditions.

\section*{Acknowledgment}
This work was sponsored by Communications Research Team (CRT), DMC R\&D Center, Samsung Electronics Co. Ltd.

\bibliographystyle{IEEEtran}
\bibliography{refs}

\begin{thebibliography}{10}
\providecommand{\url}[1]{#1}
\csname url@samestyle\endcsname
\providecommand{\newblock}{\relax}
\providecommand{\bibinfo}[2]{#2}
\providecommand{\BIBentrySTDinterwordspacing}{\spaceskip=0pt\relax}
\providecommand{\BIBentryALTinterwordstretchfactor}{4}
\providecommand{\BIBentryALTinterwordspacing}{\spaceskip=\fontdimen2\font plus
\BIBentryALTinterwordstretchfactor\fontdimen3\font minus
  \fontdimen4\font\relax}
\providecommand{\BIBforeignlanguage}[2]{{%
\expandafter\ifx\csname l@#1\endcsname\relax
\typeout{** WARNING: IEEEtran.bst: No hyphenation pattern has been}%
\typeout{** loaded for the language `#1'. Using the pattern for}%
\typeout{** the default language instead.}%
\else
\language=\csname l@#1\endcsname
\fi
#2}}
\providecommand{\BIBdecl}{\relax}
\BIBdecl

\bibitem{massive_mimo5}
F.~Rusek, D.~Persson, B.~K. Lau, E.~G. Larsson, T.~L. Marzetta, O.~Edfors, and
  F.~Tufvesson, ``Scaling up {MIMO}: Opportunities and challenges with very
  large arrays,'' \emph{IEEE Signal Processing Magazine}, vol.~30, no.~1, pp.
  40--60, Jan. 2013.

\bibitem{fdmimo}
Y.~Nam, B.~L. Ng, K.~Sayana, Y.~Li, J.~Zhang, Y.~Kim, and J.~Lee,
  ``Full-dimension {MIMO} ({FD-MIMO}) for next generation cellular
  technology,'' \emph{IEEE Communications Magazine}, vol.~51, no.~6, pp.
  172--179, Jun. 2013.

\bibitem{massive_mimo6}
E.~G. Larsson, O.~Edfors, F.~Tufvesson, and T.~L. Marzetta, ``Massive {MIMO}
  for next generation wireless systems,'' \emph{IEEE Communications Magazine},
  vol.~52, no.~2, pp. 186--195, Feb. 2014.

\bibitem{massive_mimo1}
T.~L. Marzetta, ``Noncooperative cellular wireless with unlimited numbers of
  base station antennas,'' \emph{IEEE Transactions on Wireless Communications},
  vol.~9, no.~11, pp. 3590--3600, Nov. 2010.

\bibitem{CK}
C.~K. Au-Yeung, D.~J. Love, and S.~Sanayei, ``Trellis coded line packing: Large
  dimensional beamforming vector quantization and feedback transmission,''
  \emph{IEEE Transactions on Wireless Communications}, vol.~10, no.~6, pp.
  1844--1853, Jun. 2011.

\bibitem{upa_codebook}
J.~Li, X.~Su, J.~Zeng, Y.~Zhao, S.~Yu, L.~Xiao, and X.~Xu, ``Codebook design
  for uniform rectangular arrays of massive antennas,'' \emph{Proceedings of
  IEEE Vehicular Technology Conference}, Jun. 2013.

\bibitem{jsdm}
A.~Adhikary, J.~Nam, J.~Ahn, and G.~Caire, ``Joint spatial division and
  multiplexing—the large-scale array regime,'' \emph{IEEE Transactions on
  Information Theory}, vol.~59, no.~10, pp. 6441--6463, Oct. 2013.

\bibitem{tcom_ntcq}
J.~Choi, Z.~Chance, D.~J. Love, and U.~Madhow, ``Noncoherent trellis coded
  quantization: A practical limited feedback technique for massive {MIMO}
  systems,'' \emph{IEEE Transactions on Communications}, vol.~61, no.~12, pp.
  5016--5029, Dec. 2013.

\bibitem{dawei}
D.~Ying, F.~W. Vook, T.~A. Thomas, D.~J. Love, and A.~Ghosh, ``Kronecker
  product correlation model and limited feedback codebook design in a 3{D}
  channel model,'' \emph{Proceedings of IEEE International Conference on
  Communications}, Jun. 2014.

\bibitem{tec}
\BIBentryALTinterwordspacing
J.~Choi, D.~J. Love, and T.~Kim, ``Trellis-extended codebooks and successive
  phase adjustment: A path from {LTE}-{A}dvanced to {FDD} massive {MIMO}
  systems,'' \emph{IEEE Transactions on Wireless Communications}, accepted for
  publication. [Online]. Available: \url{http://arxiv.org/abs/1402.6794}
\BIBentrySTDinterwordspacing

\bibitem{song}
S.~Noh, M.~D. Zoltowski, Y.~Sung, and D.~J. Love, ``Pilot beam pattern design
  for channel estimation in massive {MIMO} systems,'' \emph{IEEE Journal of
  Selected Topics in Signal Processing}, vol.~8, no.~5, pp. 787--801, Oct.
  2014.

\bibitem{cl_training_jstsp}
J.~Choi, D.~J. Love, and P.~Bidigare, ``Downlink training techniques for {FDD}
  massive {MIMO} systems: Open-loop and closed-loop training with memory,''
  \emph{IEEE Journal of Selected Topics in Signal Processing}, vol.~8, no.~5,
  pp. 802--814, Oct. 2014.

\bibitem{3dmimo1}
B.~L. Ng, Y.~Kim, J.~Lee, Y.~Li, Y.~Nam, J.~Zhang, and K.~Sayana, ``Fulfilling
  the promise of massive {MIMO} with {2D} active antenna array,''
  \emph{Globecom Workshops}, Dec. 2012.

\bibitem{3dmimo2}
T.~A. Thomas and F.~W. Vook, ``Transparent user-specific {3D} {MIMO} in {FDD}
  using beamspace methods,'' \emph{Proceedings of IEEE Global
  Telecommunications Conference}, Dec. 2012.

\bibitem{3dmimo3}
Y.~Li, X.~Ji, D.~Liang, and Y.~Li, ``Dynamic beamforming for three-dimensional
  {MIMO} technique in {LTE}-{A}dvanced networks,'' \emph{International Journal
  of Antennas and Propagation}, 2013.

\bibitem{lte}
\emph{Evolved universal terrestrial radio access (E-UTRA): physical channels
  and modulation}, 3GPP TS 36.211 v11.0.0 Std., Sep. 2012.

\bibitem{zfbf}
T.~Yoo and A.~Goldsmith, ``On the optimality of multiantenna broadcast
  scheduling using zero-forcing beamforming,'' \emph{IEEE Journal on Selected
  Areas in Communications}, vol.~24, no.~3, pp. 528--541, Mar. 2006.

\bibitem{tm_correlated1}
B.~Banister and J.~Zeidler, ``Feedback assisted transmission subspace tracking
  for {M}{I}{M}{O} systems,'' \emph{IEEE Journal on Selected Areas in
  Communications}, vol.~21, pp. 452--463, May 2003.

\bibitem{tm_correlated2}
J.~Yang and D.~Williams, ``Transmission subspace tracking for {M}{I}{M}{O}
  systems with low-rate feedback,'' \emph{IEEE Transactions on Communications},
  vol.~55, no.~8, pp. 1629--1639, Aug. 2007.

\bibitem{tm_correlated8}
K.~Huang, R.~W. {Heath Jr.}, and J.~G. Andrews, ``Limited feedback beamforming
  over temporally correlated channels,'' \emph{IEEE Transaction on Signal
  Processing}, vol.~57, no.~5, pp. 1959--1975, May 2009.

\bibitem{tm_correlated4}
T.~Kim, D.~J. Love, and B.~Clerckx, ``{MIMO} system with limited rate
  differential feedback in slow varying channel,'' \emph{IEEE Transactions on
  Communications}, vol.~59, no.~4, pp. 1175--1180, Apr. 2010.

\bibitem{tm_correlated5}
J.~Choi, B.~Clerckx, N.~Lee, and G.~Kim, ``A new design of polar-cap
  differential codebook for temporally/spatially correlated {MISO} channels,''
  \emph{IEEE Transactions on Wireless Communications}, vol.~11, no.~2, pp.
  703--711, Feb. 2012.

\bibitem{tm_correlated6}
J.~Choi, B.~Clerckx, and D.~J. Love, ``Differential codebook for general
  rotated dual-polarized {MISO} channels,'' \emph{Proceedings of IEEE Global
  Telecommunications Conference}, Dec. 2012.

\bibitem{tm_correlated9}
J.~Mirza, P.~A. Dmochowski, P.~J. Smith, and M.~Shafi, ``A differential
  codebook with adaptive scaling for limited feedback {MU} {MISO} systems,''
  \emph{IEEE Wireless Communications Letters}, vol.~3, no.~1, pp. 2--5, Feb.
  2014.

\bibitem{alu}
{Alcatel-Lucent Shanghai Bell}, ``Considerations on {CSI} feedback enhancements
  for high-priority antenna configurations,'' \emph{3GPP TSG RAN WG1 \#66,
  R1-112420}, Aug. 2011.

\bibitem{prok}
J.~G. Proakis, \emph{Digital Communication}, 4th~ed.\hskip 1em plus 0.5em minus
  0.4em\relax New York: McGraw-Hill, 2000.

\end{thebibliography}

\end{document}